# Social versus Moral preferences in the Ultimatum Game: A theoretical model and an experiment

Valerio Capraro

Department of Economics, Middlesex University London, United Kingdom
V.Capraro@mdx.ac.uk

**Abstract**

In the Ultimatum Game (UG) one player, named "proposer", has to decide how to allocate a certain amount of money between herself and a "responder". If the offer is greater than or equal to the responder's minimum acceptable offer (MAO), then the money is split as proposed, otherwise, neither the proposer nor the responder get anything. The UG has intrigued generations of behavioral scientists because people in experiments blatantly violate the equilibrium predictions that self-interested proposers offer the minimum available non-zero amount, and self-interested responders accept. Why are these predictions violated? Previous research has mainly focused on the role of social preferences. Little is known about the role of general moral preferences for doing the right thing, preferences that have been shown to play a major role in other social interactions (e.g., Dictator Game and Prisoner's Dilemma). Here I develop a theoretical model and an experiment designed to pit social preferences against moral preferences. I find that, although people recognize that offering half and rejecting low offers are the morally right things to do, moral preferences have no causal impact on UG behavior. The experimental data are indeed well fit by a model according to which: (i) high UG offers are motivated by inequity aversion and, to a lesser extent, self-interest; (ii) high MAOs are motivated by inequity aversion.

# 1. Introduction

The Ultimatum Game (UG) has intrigued generations of social scientists. Thirty-five years after its invention, searching "ultimatum game" on google scholar generates over 58.000 results, and the 1982 paper by Güth, Schmittberger and Schwarze, which famously introduced the game, has collected over 4.000 citations. The UG is, without any doubt, among the most studied economic games in social and biological sciences. But why?

One factor is simplicity. The UG can be explained to literally everyone. There are two players and a pie. The first player, named "proposer", has to decide how to divide the pie between herself and the second player, named "responder". At the same time, the responder decides his minimum acceptable offer (MAO). If the proposer's offer is higher than or equal to the responder's MAO, then the pie is divided between the proposer and the responder as agreed; otherwise, no one get anything.

Also the equilibrium analysis is very simple, at least if one assumes, as neo-classical economic theory does, that both the proposer and the responder are motivated by self-interest, and that this is common knowledge: a self-interested responder would accept any non-zero offer; anticipating this behavior, a self-interested proposer would offer the smallest positive amount possible.

However, these predictions are blatantly violated in laboratory experiments: offers below 25% are typically rejected, and the vast majority of offers lie between 30% and 50% (Camerer, 2003; Fehr & Schmidt, 1999; Fehr & Fischbacher, 2003; Güth, Schmittberger and Schwarze, 1982; Güth & Kocher, 2014; Henrich et al, 2001). Why do people violate the equilibrium predictions?

Here it starts the interesting part of the story. As Güth and Kocher say in their exceptional review of thirty years of UG experiments: "motivations behind decisions in the ultimatum game are diverse" (Güth & Kocher, 2014, p. 397), and "[w]hat could not easily be anticipated is how socially, motivationally, and emotionally rich already the very simple ultimatum game turned out to be" (ibid., p. 406).

What are these motivations? I will review them in detail in the next section. In short, previous research has mainly focused on the role of social preferences concluding that: high offers are driven by a combination of inequity aversion and self-interest[1] (e.g., Camerer, 2003); rejections of low offers are driven by a combination of inequity aversion and spitefulness (e.g., Yamagishi et al 2012; Brañas-Garza, Espín, Exadaktylos & Hermann, 2014).

However, in the last years, several researchers have provided evidence that people are not solely motivated by outcome-based social preferences: they are also motivated by general preferences for doing the morally right thing. Capraro and Rand (2018) found that altruistic behavior in the Dictator Game and cooperative behavior in the Prisoner's Dilemma are partly driven by non-outcome-based preferences for doing the "nice" thing. Krupka and Weber (2013) found that framing effects in the Dictator

---

[1]Throughout the whole paper, I will consider "self-interest" as a degenerate social preference, having in mind that self-interest can be recovered from social preferences assuming that the weight on the others' payoffs is equal to 0.

Game are due to changes in the perception of what is the morally appropriate thing to do. Numerous theoretical models have also been advanced to formalize people's tendency to do the morally right thing (Alger & Weibull, 2013; Brekke, Kverndokk & Nyborg, 2003; DellaVigna, List & Malmendier, 2012; Huck, Kübler, & Weibull, 2012; Kessler & Leider, 2012; Kimbrough & Vostroktunov, 2016; Krupka & Weber, 2013; Lazear, Malmendier & Weber, 2012; Levitt & List, 2017; Weber, Kopelman & Messick, 2004).

These observations raise important questions regarding the Ultimatum Game. Do moral preferences play a role in the Ultimatum Game? Could it be the case that proposers make high offers just because they think that this is the right thing to do? Could it be the case that responders reject low offers because they think that this is the right thing to do?

Here, I would like to shed light on these questions. The paper is structured as follows. Section 2 reviews previous literature and finds that, indeed, very little is known about the role of moral preferences on UG behavior. Section 3 reports a preliminary study which shows that people self-report that making high offers and rejecting low offers are the morally right things to do. Section 4 develops the theoretical background towards my main experiment, the aim of which is to investigate the extent to which making high offers and rejecting low offers are *caused* by moral preferences. Section 5 reports the main experiment and finds that moral preferences play no major roles in determining UG behavior. Instead, the best explanation for the results is that: (i) high UG offers are motivated by inequity aversion and, to a lesser extent, self-interest; (ii) rejection of low offers is motivated by inequity aversion. Section 6 concludes.

## 2. Literature review

The literature on the UG is enormous, counting thousands, if not tens of thousands, papers. In this section I provide a short, but comprehensive, description of the main frameworks that have been advanced to explain the behavior of proposers and responders.

I will mainly focus on theories grounded on outcome-based social preferences: *inequity aversion*, a desire to decrease payoff differences; *social efficiency*, a desire to increase the sum of the payoffs of all players; *spitefulness*, a desire to be better off than the other player(s), even at one's own cost; and *self-interest*, a desire to increase one's own payoff (which can be thought as a degenerate social preference, where subjects care about others' payoffs with weight equal to 0). I will focus on these motivations because previous research has found compelling evidence that they all play a major role in driving people's decision making (Bolton & Ockenfels, 2000; Capraro, Corgnet, Espín & Hernán-González, 2017; Charness & Rabin, 2002; Corgnet, Espín & Hernán-González, 2015; Engelmann & Strobel, 2004; Fehr & Schmidt, 1999; Fehr & Schmidt, 2006; Messick & McClintock, 1968; Van Lange, Otten, DeBruin & Joireman, 1997).

Of course, since my goal is to investigate the extent to which moral preferences determine UG behavior, I will also discuss studies exploring the effect of moral preferences. To the best of my knowledge, there are only two such studies

(Kimbrough & Vostroknutov, 2016; Eriksson, Strimling, Anderson & Lindholm, 2017).

Finally, in the last paragraph of this section, I will also shortly discuss approaches beyond social and moral preferences.

*Responders*

It has been widely observed that responders reject low offers and that fair offers are virtually always accepted (Güth et al, 1982; Camerer, 2003; Güth & Kocher, 2014). Somewhat more surprisingly, it has been noted that sometimes people even reject super-fair offers, larger than half of the pie (Güth, Schmidt & Sutter, 2003). Why so?

Earlier accounts took these results as evidence that responders are motivated by inequity aversion: they reject low offers because they create inequity between the proposer and the responder (Fehr & Schmidt, 1999; Bolton & Ockenfels, 2000). By assuming that some responders are averse to advantageous inequalities these accounts can easily explain also why experimenters see a small proportion of rejection of super-fair offers.

However, this interpretation has been recently found incomplete, as evidence emerged of a class of subjects who reject unfair offers motivated by spiteful preferences, that is, by a desire to be better off than the proposer, motivation, which, in the case of a UG responder who receives a low offer, happens to generate the same behavior as inequity aversion (Yamagishi et al 2012; Brañas-Garza, Espín, Exadaktylos & Hermann, 2014; Espín, Exadaktylos, Hermann & Brañas-Garza, 2015; Yamagishi et al, 2017). For example, Brañas-Garza et al (2014) demonstrated that there are subjects who reject low offers in the UG but donate nothing when playing the Dictator Game (DG) as dictators (the DG is similar to the UG, but the responder makes no choice and only gets what the first person, the dictator, decides to give. The DG allows to explore the effect of inequity aversion without any strategic confound. Thus, if rejections of low offers were solely motivated by inequity aversion, then people who reject low offers in the UG should give in the DG).

Another stream of research investigated the effect of time pressure and time delay on rejection rates. Grimm & Mengel (2011) found that an exogenously imposed delay decreases rejection of low offers. Complementarily, Sutter et al (2013) found that time pressure increases rejection of low offers. Since time pressure increases both inequity aversion and spitefulness (Capraro et al, 2017), these results fit well with the aforementioned interpretation that rejection of low offers is driven by a combination of inequity aversion and spitefulness.

Neuroscientific research has shed further light on the topic by suggesting a role for both emotion and cognition in the rejection of low offers (Crockett, Clark, Tabibnia, Lieberman & Robbins, 2008; Sanfey, Rilling, Aronson, Nystrom & Cohen, 2003; van't Wout, Kahn, Sanfey & Aleman, 2006; Yamagishi, Horita, Takagishi, Shinada, Tanida & Cook, 2009). For example, Yamagishi et al (2017) concluded that "[s]piteful punishers anticipate pleasure from seeing their target suffer by rejecting unfair offers, while altruistic punishers reject unfair offers as a deliberate attempt to punish norm violators (as in the TPPG [Third-Party Punishment Game]), rather than seeking pleasure from enforcing fairness norms". How to interpret this conclusion from the perspective of social (and moral) preferences is unclear. It seems to suggest that altruistic punishers may not be guided solely by a desire to decrease inequality, but also (and potentially only) by a more abstract desire of doing the morally right thing.

To the best of my knowledge, only two studies explored directly a potential effect of morality on rejection of low offers. Kimbrough & Vostroknutov (2016) found a positive correlation between rejection of low offers and rule-following in an unrelated task. Although interesting, this correlational study does not clarify whether rejections of low offers are actually caused by preferences for following the rules. An exogenous manipulation to show causality has been implemented by Eriksson et al (2017), who found that when the action of rejecting is framed as "payoff reduction" (of the other participant's payoff), then people rejection rates decline, and the action of "reducing the payoff" is judged to be more morally wrong, compared to when the choice is simply called "rejection". This provides a first piece of evidence that moral considerations play a causal role in determining rejection of low offers. However, how big this role is remains unclear. Is rejection of low offers entirely driven by moral preferences? What is the relative proportion of inequity aversion and moral preferences in determining rejections of low offers?

None of the previous studies can answer these questions. Not even those providing evidence of a correlation between rejection of low offers and DG giving (Branas-Garza et al, 2014; Capraro et al, 2015). The reason is simple: DG giving is not evidence of inequity aversion as it is partly motivated by general moral preferences for doing the right thing (Capraro & Rand, 2018).

To summarize this part, previous research suggests that rejection of low offers is driven by a combination of spitefulness and fairness enforcement. However, it is not clear the extent to which fairness enforcement is driven by outcome-based preferences for minimizing inequities or by a general moral preference for doing the morally right thing.

*Proposers*

Proposers offer typically between 30% and 50% of the pie, well above the equilibrium predictions (Güth et al, 1982; Camerer, 2003; Güth & Kocher, 2014). Why so?

One popular explanation assumes that, actually, only responders violate the theory of self-interest: assuming that proposers know that low offers will be rejected, then it is in their self-interest to make high offers (Chen et al, 2017; Roth, Prasnikar, Okuno-Fujiwara & Zamir,1991; Wells & Rand, 2013). For example, Roth et al (1991) reported UG experiments from four countries showing that the most common offer was the choice maximizing the proposer's payoff, conditional on the rejections rate. However, this explanation has been criticized by more recent accounts. Lin & Sunder (2002) found that the 60% of the proposers offered more than the payoff-maximizing amount. Henrich et al (2001) analyzed UG offers in 15 small-scale societies and found that, in most cases, the modal offer was substantially higher than the payoff-maximizing one. This suggests that self-interest alone is unlikely to explain proposers' behavior. Following this line, Camerer (2003) concluded that "some of their generosity in ultimatum games is altruistic rather than strategic" (p. 56).

Is this altruistic component driven by inequity aversion or by a general moral tendency of doing the morally right thing? To the best of my knowledge, only one study has explored the role of moral preferences on UG offers. Kimbrough & Vostroknutov (2016) found no correlation between UG offers and rule-following in a separated task. This provides a first piece of evidence that moral preferences have little to do with UG offers.

To summarize this part, previous research suggests that high offers in the UG are driven by a combination of self-interest and fairness/altruism. However, very little is known about whether fairness/altruism is driven by outcome-based social preferences for minimizing inequities of by moral preferences for doing the morally right thing.

*Other explanations*

Theories beyond social and moral preferences have also been proposed. For example, Capraro et al (2015) applied the notion of cooperative equilibrium (Capraro, 2013; Barcelo & Capraro, 2015) to the UG and found that it makes reasonable predictions of average behavior of proposers. Similarly, Schuster (2017) and Suleiman (2017) introduced novel solution concepts providing theoretical explanations for why UG offers are, on average, typically close to the Golden Ratio. However, although these models make reasonably good predictions of proposers' average behavior, there is no evidence that real people actually follow any of the supposed lines of reasoning.

Other frameworks make use of bounded rationality and explain the behavior of proposers as a failure to use backward induction (Brenner & Vriend, 2006). However, experimental research has shown that bounded rationality is likely to play very little role in determining proposers' decisions. For example, Avrahami et al (2010) conducted a repeated ultimatum game with random re-matching and found a very quick convergence to proposers offering and responders demanding the equal share. Similarly, and even more convincingly, Cooper and Dutcher (2011), in a meta-study, provided evidence that behavior in repeated ultimatum games with random re-matching after each interaction seems similar to learning a norm: with experience, acceptance rates of low offers even decline.

Another line of research explores the evolutionary mechanisms that can promote fairness in a population of selfish individuals. For example, Nowak, Page & Sigmund (2000) showed that fairness will evolve if the proposer can obtain information about responder's past rejection rates. Szolnoki, Perc and Szabó (2012a) found that accuracy in strategy imitations promotes the evolution of fairness. Szolnoki, Perc and Szabó (2012b) found that empathy leads to the evolution of equality. Rand et al (2013) applied stochastic evolutionary game theory, where agents make mistakes when judging the payoffs and strategies of others, to show that natural selection favors fairness. Although these and similar studies are extremely interesting because they may help us understand the ultimate origin of fairness, the focus of the current work is on proximate explanations.

To summarize this literature review section, very little is known about the role of moral preferences in UG behavior. The goal of the next sections is to shed some light on this role.

### 3. A preliminary study: What's the right thing to do in the UG?

One preliminary question is in order. Do people recognize that the morally right thing to do in the UG is to give half and to reject low offers?

To answer these questions, I conducted a small study in which subjects (living in the US and recruited using the online labor market Amazon Mechanical Turk, AMT[2]) were presented the rules of the UG and then were asked the following questions in random order: "What do you think is the morally right offer?" and "What do you think it is the morally right minimum offer for you to accept?" I refer to the Appendix for the exact instructions[3].

Results showed that proposers overwhelmingly said that offering half was the morally right thing to do (N=177, 92% chose "offer half"), while responders that demanding half was the morally right thing to do (N=176, 72% chose "demand half").

These results confirm that subjects recognize that the Ultimatum Game has a moral component. However, they do not clarify whether this moral component plays a causal role in determining UG behavior. Answering this question is the goal of my main study.

### 4. Theoretical model and behavioral predictions

Having established, by the literature review, that previous research is essentially silent about the role of morality on UG behavior, and having established, by a preliminary study, that subjects think that the UG has a moral component, here I develop a theoretical background to test the hypothesis that morality has a causal impact on UG behavior.

*The Trade-Off game*

In Capraro & Rand (2018) we used the Trade-Off Game (TOG) to show that a significant proportion of subjects are motivated by a general morality preference, and that this morality preference plays a major role in determining Dictator game and Prisoner's dilemma behavior.

---

[2]Experiments on AMT are easy to implement and fast and cheap to realize, because subjects participate from their homes in an incentivized survey that takes no more than a few minutes. This allows researcher to decrease the participation fee and the game stakes significantly, without damaging the quality of the results. Numerous studies using economic games have indeed found that data gathered on AMT are of no less quality than those collected using physical laboratories (Amir, Rand, & Gal, 2012; Arechar, Gächter & Molleman, 2018; Brañas-Garza, Capraro, Rascón-Ramírez, 2018; Horton, Rand, & Zeckhauser, 2011; Paolacci, Chandler & Ipeirotis, 2010; Paolacci & Chandler, 2014; Mason & Suri, 2012; Thomas & Clifford, 2017).

[3]In the original study, subjects were asked to report what they think is the morally right thing to do also in the Dictator Game, the Prisoner's Dilemma and the Trade-Off game. Results were reported elsewhere (Capraro and Rand, 2018). Here I focus on the Ultimatum Game.

In the TOG, the decision-maker has to decide between two allocations of money that affect three players, the decision maker herself and two other players. Crucially, one allocation minimizes inequities: E = [13 13 13] (i.e., 13 cents to each of the three players); while the other one maximizes the social welfare: S = [15 23 13] (e.g., the decision maker gets 15 cents, the second player gets 23 cents, the third player gets 13 cents). The TOG is played in either of two frames. In the *Give* frame, choices are presented in such a way that participants are led to believe that the socially efficient choice is the right thing to do (e.g., the allocation [15 23 13] is named "nice choice", while the allocation [13 13 13] is called "not nice choice"); in the *Equalize* frame, the choices are presented in such a way that participants are led to believe that the equitable choice is the right thing to do (e.g., the allocation [13 13 13] is named "nice choice", while the allocation [15 23 13] is called "not nice choice". The idea is that inequity averse and spiteful subjects will choose the allocation [13 13 13] regardless of the frame; socially efficient and self-interested subjects will choose the allocation [15 23 13] regardless of the frame; subject who are motivated by general preferences for doing the right thing will choose the option that is presented as the nice option, regardless of the frame[4]. In Capraro & Rand (2018) we showed two main results: (i) the frame has a massive impact on choices; (ii) people who make the nice choice in TOG are more likely to act altruistically in a one-shot Dictator Game (DG) and cooperatively in a one-shot Prisoner's Dilemma (PD). This led us to conclude that DG altruism and PD cooperation are partly driven by a desire of doing the right thing to do.

Here I apply a similar technique: participants will play both the TOG and the UG and I will explore the correlation between framing manipulation in the TOG and UG behavior. Comparing the results with the formal predictions that I will derive in the next section will allow me to deduce which motivations are responsible for UG behavior.

*Theoretical model*

*Notation*

- N will denote the frame-dependent nice choice in the TOG, that is, the choice that is presented to be the nice choice (remember that E and S denote respectively the equitable and the social efficient choices).
- $p_E$ will denote the offer made by a UG proposer chosen at random among those who choose option E in the TOG. Similarly, I define $p_S$ and $p_N$. And similarly I define $r_E$, $r_S$, and $r_M$ (where r stands for responder).
- Thus, $p_E$, $p_S$, etc., are discrete random variables with values in the interval [0,1]. I will denote $M(p_E)$, $M(p_S)$, etc., their mean value, which I assume to be well defined and exact (i.e., the mean value that one would find, had he or she the possibility to collect countably many observations).

---

[4]Of course, framing effect in this case can be also driven by an experimenter demand effect. This is not an issue, as susceptibility to experimenter demand effects is an instance of norm-sensitivity (Kimbrough & Vostroknutov, 2016).

*Model*

I assume that each subject comes with a motivation (or preference, or type) *m* and chooses in UG and TOG according to *m*, that is, motivations are constant across games[5]. Moreover, I assume that there are only three possible motivations: inequity aversion (IA), self-interest (SI), and morality (M)[6]. I make very light assumptions on the distribution of these motivations in the population. For the formal arguments to work, it is enough that a significant proportion of the population is motivated by each of these motivations. In formal terms, I introduce a parameter α representing the proportion of IA versus M subjects and a parameter β representing the proportion of SI versus M subjects. With this notation, I require that α, β > 0.

In the model, subjects make a decision in both the TOG and the UG. Since the TOG has no strategic component, I assume that the decision in the TOG is entirely dependent on the subject's type as follows: IA subjects choose E regardless of the TOG frame; SI subjects choose S, regardless of the TOG frame; M subjects choose N, regardless of the TOG frame.

Since the UG has a strategic component, I cannot make a priori assumptions on how subjects will play the UG as a function of their motivations. Indeed, the whole point of the model is to derive predictions about the correlation between TOG and UG behaviors for all possible assumptions about the driving motivations of UG behavior.

Finally, two minor and essentially only technical assumptions. I assume that $M(p_{SI}) < M(p_{IA})$, that is, that the average offer made by IA subjects is significantly smaller than the average offer made by SI subjects. Two observations about this requirement: first, it allows to reduce the number of cases to be analyzed (in the next subsection), using a reasonable assumption that has been supported by experimental studies showing a positive correlation between DG giving and UG offers (Branas-Garza et al, 2014; Capraro et al, 2015); second, the derivation of predictions for the remaining cases $M(p_{SI}) = M(p_{IA})$ and $M(p_{SI}) > M(p_{IA})$ is very simple, as it is simple to the see that they are not satisfied by the results of the experiment. For similar reasons, I assume that $M(r_{SI}) < M(r_{IA})$.

*Behavioral predictions*

The general structure of this section is as follows. For every order position of $M(p_M)$ with respect to $M(p_{SI})$ and $M(p_{IA})$, I will deduce, when possible, an order relation between $M(p_E)$ and $M(p_S)$. Similarly, for responders. Testing which of these order relations are satisfied will be the goal of the main experiment.

---

[5] The assumption that motivations/preferences are constant across games is somehow implicit in behavioral economics, and has received some initial support from research showing that prosocial choices in different economic games are all correlated (Capraro, Jordan & Rand, 2014; Capraro, Smyth, Mylona & Niblo, 2014; Peysakhovich, Nowak & Rand 2014; Reigstad, Strømland & Tinghög, 2017), which led to the definition of "cooperative phenotype" (Peysakhovich, Nowak & Rand 2014).

[6] My design will not allow me to distinguish spitefulness from inequity aversions and social efficiency from self-interested. Although, I acknowledge, this is a general limitation of the design, I believe that this is not an issue, because the experiment is designed to distinguish social preferences from moral preferences. Understanding which social preference is at play is not the purpose of this work. Although I will be able to partly answer this question.

*Case 1. If $M(p_M) < M(p_{SI}) < M(p_{IA})$*

Consider the TOG in the Equalize frame. The equitable choice E is taken by IA subjects and by M subjects, thus: $M(p_E) = \alpha M(p_{IA}) + (1-\alpha)M(p_M)$. The efficient choice S is taken by SI subjects only, thus: $M(p_S) = M(p_{SI})$. Since $M(p_{SI})$ is strictly between $M(p_{IA})$ and $M(p_M)$, and since $\alpha > 0$, I cannot conclude anything about whether $M(p_E)$ is smaller or greater than $M(p_S)$. It can be anything.

Consider the TOG in the Give frame. The equitable choice E is taken by IA subjects only, thus: $M(p_E) = M(p_{IA})$. The efficient choice S is taken by SI subjects and by M subjects, thus: $M(p_S) = \beta M(p_{SI}) + (1-\beta)M(p_M)$. Since $M(p_{SI})$ and $M(p_M)$ are both strictly smaller than $M(p_{IA})$, so it will any of their convex combinations. Thus: $M(p_S) < M(p_E)$.

*Prediction 1.* If $M(p_M) < M(p_{SI}) < M(p_{IA})$, then, in the Give frame we should have $M(p_S) < M(p_E)$, while no particular order relation between $M(p_S)$ and $M(p_E)$ is predicted in the Equalize frame.

*Case 2. If $M(p_M) = M(p_{SI}) < M(p_{IA})$*

Consider the TOG in the Equalize frame. The equitable choice E is taken by IA subjects and by M subjects, thus: $M(p_E) = \alpha M(p_{IA}) + (1-\alpha)M(p_M)$. The efficient choice S is taken by SI subjects only, thus: $M(p_S) = M(p_{SI}) = M(p_M)$. Since $\alpha > 0$ and $M(p_{IA}) > M(p_M)$, then: $M(p_S) < M(p_E)$.

Consider the TOG in the Give frame. The equitable choice E is taken by IA subjects only, thus: $M(p_E) = M(p_{IA})$. The efficient choice S is taken by SI subjects and by M subjects, thus: $M(p_S) = \beta M(p_{SI}) + (1-\beta)M(p_M)$. Thus, $M(p_S) < M(p_E)$.

*Prediction 2.* If $M(p_M) = M(p_{SI}) < M(p_{IA})$, then $M(p_S) < M(p_E)$ in both TOG frames.

*Case 3. If $M(p_{SI}) < M(p_M) < M(p_{IA})$*

Consider the TOG in the Equalize frame. The equitable choice E is taken by IA subjects and by M subjects, thus: $M(p_E) = \alpha M(p_{IA}) + (1-\alpha)M(p_M)$. The efficient choice S is taken by SI subjects only, thus: $M(p_S) = M(p_{SI})$. Since $M(p_{SI}) < M(p_M), M(p_{IA})$, then: $M(p_S) < M(p_E)$.

Consider the TOG in the Give frame. The equitable choice E is taken by IA subjects only, thus: $M(p_E) = M(p_{IA})$. The efficient choice S is taken by SI subjects and by M subjects, thus: $M(p_S) = \beta M(p_{SI}) + (1-\beta)M(p_M)$. Then, also in this case: $M(p_S) < M(p_E)$.

*Prediction 3.* If $M(p_{SI}) < M(p_M) < M(p_{IA})$, then $M(p_S) < M(p_E)$ in both TOG frames.

*Case 4. If $M(p_{SI}) < M(p_M) = M(p_{IA})$*

Consider the TOG in the Equalize frame. The equitable choice E is taken by IA subjects and by M subjects, thus: $M(p_E) = \alpha M(p_{IA}) + (1-\alpha)M(p_M)$. The efficient choice S is taken by SI subjects only, thus: $M(p_S) = M(p_{SI})$. Then: $M(p_S) < M(p_E)$.

Consider the TOG in the Give frame. The equitable choice E is taken by IA subjects only, thus: $M(p_E) = M(p_{IA})$. The efficient choice S is taken by SI subjects and by M subjects, thus: $M(p_S) = \beta M(p_{SI}) + (1-\beta)M(p_M)$. Then: $M(p_S) < M(p_E)$.

*Prediction 4.* If $M(p_{SI}) < M(p_M) = M(p_{IA})$, then $M(p_S) < M(p_E)$ in both TOG frames.

*Case 5. If $M(p_{SI}) < M(p_{IA}) < M(p_M)$*

Consider the TOG in the Equalize frame. The equitable choice E is taken by IA subjects and by M subjects, thus: $M(p_E) = \alpha M(p_{IA}) + (1-\alpha)M(p_M)$. The efficient choice S is taken by SI subjects only, thus: $M(p_S) = M(p_{SI})$. Then: $M(p_S) < M(p_E)$.

Consider the TOG in the Give frame. The equitable choice E is taken by IA subjects only, thus: $M(p_E) = M(p_{IA})$. The efficient choice S is taken by SI subjects and by M subjects, thus: $M(p_S) = \beta M(p_{SI}) + (1-\beta)M(p_M)$. Since $M(p_{IA})$ is between $M(p_{SI})$ and $M(p_M)$ and since $\beta > 0$, we cannot deduce predictions regarding the order relation between $M(p_S)$ and $M(p_E)$.

*Prediction 5.* If $M(p_{SI}) < M(p_{IA}) < M(p_M)$, then $M(p_S) < M(p_E)$ in the Equalize frame, while in the Give frame there is no special order relation.

Of course, the same predictions hold also for the responder's behavior.

## 5. Main experiment

*Subjects*

Subjects, living in the US at the time of the experiment, were recruited using the online labor market Amazon Mechanical Turk (AMT).

*Procedure*

Subjects were randomly assigned to one of four conditions: UG1-Give-UG2, UG2-Give-UG1, UG1-Equalize-UG2, and UG2-Equalize-UG1 (The notation X-Y-Z means that the subjects first played game X, then game Y, and finally game Z). In the UG1 subjects played the UG as proposers. In the UG2, they played the UG as responders. Proposers were given $0.10 and were asked how much, if any, they wanted to offer to another anonymous participant, who started given nothing. Responders were asked to decide their MAO. If the proposer's offer was larger than or equal to the responder's MAO, then the $0.10 was split between the two subjects according to the agreed offer. Otherwise, both subjects earned nothing. All subjects played in both roles (denoted UG1 and UG2). Between the two decisions, they played the TOG in either the Give or the Equalize frame. Standard comprehension questions were asked to make sure that subjects understood the instructions. Subjects failing any comprehension question were automatically excluded from the survey. After the data were collected, bonuses were computed and paid, on top of the participation fee ($0.50). To compute the bonuses, I randomly selected one of the two UG roles for each of the subjects. No deception was used. I refer to the Appendix for experimental instructions.

*Results*

A total of 575 subjects passed the comprehension questions. I start analyzing UG offers.

As a first step of the analysis, I show that the order of roles in which the UG is played does not matter. To this end, I conduct linear regression predicting *UG_offer* as a function of three dummy variables (*Give_frame* = 1 if a subject participated in the Give frame, and 0 if, otherwise, they participated in the Equalize frame; *Equal_choice* = 1 if a subject chose the equitable option in the TOG, and 0 if they chose the efficient option; and the variable *UG1_first* = 1 if a subject played the UG1 before the TOG, and 0 if, otherwise, they played UG1 after the TOG) and all their interactions. I find that the *UG1_first* does not significantly predict *UG_offer* (coeff = -0.011, p = 0.473), does not significantly interact with *Give_frame* (coeff = 0.002, p = 0.925), does not interact with *Equal_choice* (coeff = -0.556, p = 0.254), and does not interact with their interaction (coeff = 0.534, p = 0.334). Thus, the order of UG play does not matter and I can collapse across orders.

I now pass to the main analysis. Linear regression predicting *UG_offer* as a function of *Equal_choice* and *Give_frame* and their interaction reveals a significant interaction between *Equal_choice* and *Give_frame* (coeff = 0.072, p = 0.009). Looking at the main effects of *Equal_choice* within each frame of the TOG, I find that, in the Equalize frame, the average offer among subjects who chose S in the TOG was 48%, while the average offer among those who chose E was 46% (rank-sum, p = 0.994). However, in the give frame, the average offer among subjects who chose S was 44%, while the average offer among those who chose E was 50% (rank sum, p = 0.001). See Figure 1.

In sum, there is a significant interaction between TOG frame and UG offers, such that UG offers are positively correlated with the equitable option in the TOG but only in the Give frame. Looking at the predictions in Section 4, I find that the only hypothesis that agrees with this result is: $M(p_M) < M(p_{SI}) < M(p_{IA})$, that is, high offers are primarily driven by inequity aversion and, to a lesser extent, by self-interest. Morality plays no role in determining high offers.

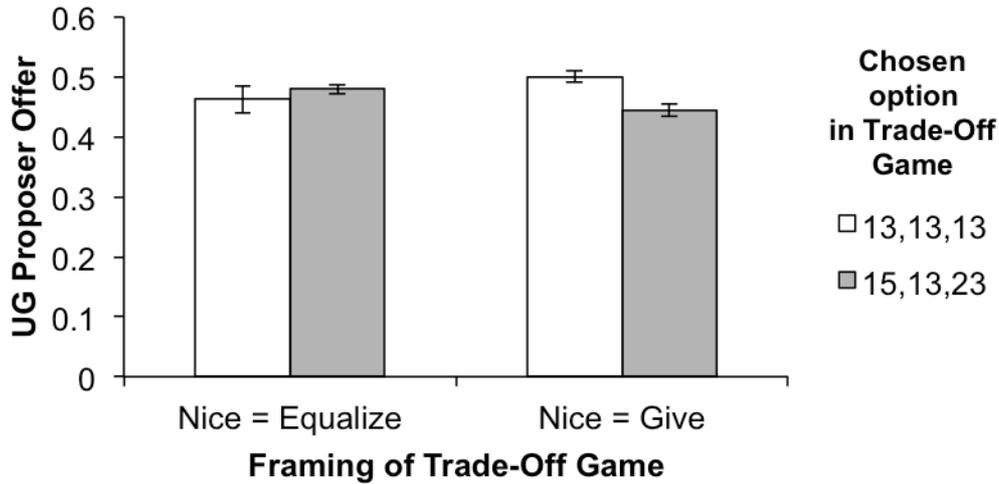

*Figure 1. Average UG offer as a function of the choice in each of the two framings of the TOG. In the Equalize frame there is no significant correlation between being nice in the TOG and UG offer. In the Give frame there is a significant correlation between "not" being nice in the TOG and UG offer.*

Next, I analyze responders' behavior. As before, as a first step of the analysis, I show that the order of play does not matter. To this end, I conduct linear regression predicting *UG_MAO* as a function of *Give_frame*, *Equal_choice*, *UG_first*, and all their interactions. I find that the *UG_first* does not significantly predict *UG_MAO* (coeff = -0.022, p = 0.295), does not significantly interact with *Give_frame* (coeff = 0.010, p = 0.784), does not significantly interact with *Equal_choice* (coeff = -0.071, p = 0.303), and with their interaction (coeff = 0.077, p = 0.329). Since the order of play does not matter, I can collapse across orders.

Now I pass to the main analysis. Linear regression predicting *UG_MAO* as a function of *Equal_choice* and *Give_frame* and their interaction reveals that there is *no* significant interaction between *Equal_choice* and *Give_frame* (coeff = -0.021, p = 0.588). However, within each TOG frame, there is a significant main effect of *Equal_choice*, such that the equitable choice is positively correlated with *UG_MAO*. More precisely, in the Equalize frame, the average *UG_MAO* among subjects who made the efficient choice in the TOG was 35%, while the average *UG_MAO* among subjects who made the equitable choice in the TOG was 41% (rank-sum, p = 0.015). In the give frame, the average *UG_MAO* among subjects who made the efficient choice in the TOG was 34%, while the average *UG_MAO* among subjects who made the equitable choice in the TOG was 38% (rank sum, p = 0.034). See Figure 2.

Thus, regardless of the TOG frame, there is a positive correlation between the equitable choice in the TOG and *UG_MAO*. Looking at the predictions in Section 4, I find that, a priori, three hypotheses satisfy these predictions (Case 2, Case 3, and Case 4). However, I can easily eliminate two of them by simply observing that people choosing S in the equalize frame are necessarily SI (thus: $M(p_S) = M(p_{SI})$), while people choosing S in the Give frame are either SI or M (thus: $M(p_E) = \beta M(p_{SI})+(1-\beta)M(p_M)$). Since the data suggest that $M(p_S)=M(p_E)$, this implies that $M(p_M)$ is likely to be equal to M($p_{SI}$) – or very close to it – i.e., Case 2 in Section 4 is the one that best describes the data.

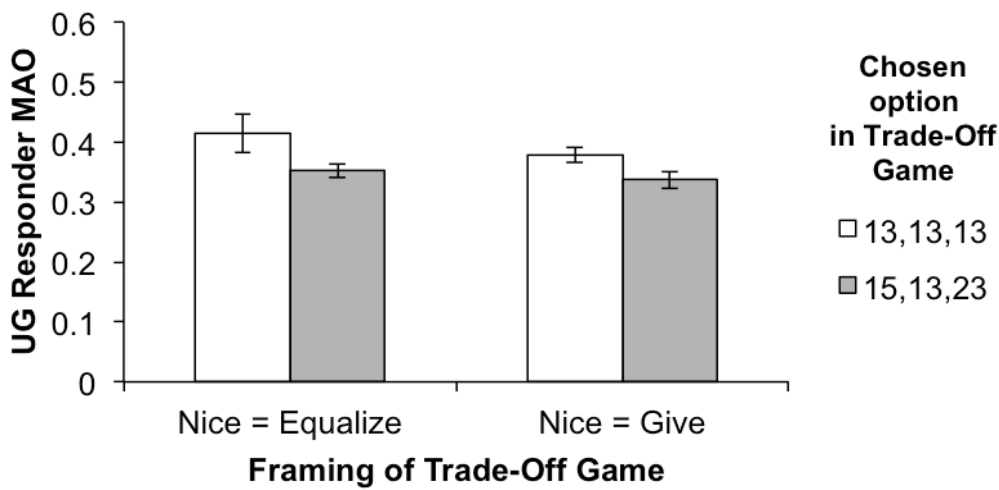

*Figure 2. Average UG MAO as a function of the choice in each of the two framings of the TOG. In both frames, there is a significant positive correlation between choosing the equitable option and the MAO (give frame: p=0.0341; equalize frame: p=0.0149).*

**General discussion**

How people play the Ultimatum Game has intrigued generations of social scientists, who have proposed a range of explanations for why experimental subjects violate the equilibrium predictions. These explanations make typically use of (a combination of) social preferences: inequity aversion, self-interest, spitefulness, and social efficiency. Despite thousands of papers, however, still very little is known about the effect of general moral preferences for doing the right thing. These preferences, in the last years, have been found to play a major role in other games involving social interactions, such as the Dictator game and the Prisoner's Dilemma (Capraro & Rand, 2018; Krupka & Weber, 2013). This raises important questions: Could it be the case that people violate the economic predictions in the UG just because they think that is the morally right thing to do? What role do moral preferences play in determining UG behavior?

The main goal of the current work was to answer these questions. I reported two experiments providing clear evidence that, despite the fact that people recognize that the morally right thing to do in the UG is to make high offers and reject low offers (preliminary study), general moral preferences do not play a major role in determining proposers' and responders' behavior (main study).

The implemented design allows me to tell also something about what motivates actual choices. I have developed a theoretical model in which people are motivated by three possible motivations (inequity aversion, self-interest, morality). I then formally derived predictions from this model and I compared the experimental results with these theoretical predictions. In doing so, I found that the results are well fit by a model assuming that: (i) high UG offers are motivated by inequity aversion and, to a lesser extent, self-interest; (ii) high MAOs are motivated by inequity aversion.

This model complements and extends the classical explanations that have been proposed to explain UG behavior by adding a moral dimension to them. Previous studies on the behavior of proposers have repeatedly found that high offers are driven by a combination of fairness and self-interest. The current results confirm this view, by adding that fairness is driven by aversion to economic inequalities, rather than by an abstract preference for doing the morally right thing. Previous results on the behavior of responders found that high MAOs are mainly driven by inequity aversion (and spitefulness). These findings confirm that inequity aversion is an important driver of high MAOs (the current design does not allow us to distinguish inequity aversion from spitefulness).

The model and the experimental design presented in this paper are silent regarding the role of spitefulness and social efficiency on UG behavior. This limitation is especially important in the case of spitefulness, as previous research has suggested that rejection of low offers is partly driven by spitefulness preferences (Yamagishi et al 2012; Brañas-Garza, Espín, Exadaktylos & Hermann, 2014; Espín, Exadaktylos, Hermann & Brañas-Garza, 2015; Yamagishi et al, 2017). In principle, also high offers might be driven by spiteful preferences, because the behavior driven by spiteful preferences happen to coincide with that driven by self-regarding preferences. However, to the best of my knowledge, no studies have explored the effect of spitefulness on UG offers. Doing a more fine-grained analysis of UG motivations by exploring also the strength of spitefulness relative to inequity aversion (for responders) and self-interest (for proposers) is certainly an important avenue for future research.

Of course, saying that morality does not play a major role in determining UG behavior is different from saying that it plays no role. The preliminary study has shown that subjects recognize that making high offers and rejecting low offers are the morally right things to do. In a similar vein, Kimbrough and Vostroknutov (2016) found a correlation between rule-following and rejection of low offers, although they found no correlation between rule-following and high offers. In any case, it is possible that special framings of the game can make the morality salient and change people's behavior accordingly. This might explain the presence of moral framings in the UG, as the ones explored reported by Eriksson et al (2017).

In a nutshell, these results shed lights on the motivations underlying behavior in the Ultimatum Game: high offers are driven by a combination of inequity aversion and self-interest; rejection of low offers are driven by inequity aversion; moral preferences play very little role in determining proposers' and responders' behavior.

# References

Alger, I., & Weibull, J. W. (2013). Homo moralis—preference evolution under incomplete information and assortative matching. *Econometrica, 81*(6), 2269-2302.

Amir, O., Rand, D. G., & Gal, Y. K. (2012). Economic Games on the Internet: The Effect of $1 Stakes. *PLoS ONE, 7*(2), e31461.

Arechar, A. A., Gächter, S., & Molleman, L. (2018). Conducting interactive experiments online. *Experimental Economics*, 21, 99-131.

Avrahami, J., Güth, W., Hertwig, R., Kareev, J., & Otsubo, H. (2013). Learning (not) to yield: An experimental study of evolving ultimatum game behavior. *The Journal of Socio-Economics*, 47, 47-54.

Barcelo, H., & Capraro, V. (2015). Group size effect on cooperation in one-shot social dilemmas. *Scientific Reports*, 5, 7937.

Bolton, G. E., & Ockenfels, A. (2000). ERC: A theory of equity, reciprocity, and competition. *The American Economic Review*, 90, 166-193.

Brañas-Garza, P., Capraro, V., Rascón-Ramírez, E. (2018). Gender differences in altruism on Mechanical Turk: Expectations and actual behaviour. *Available at SSRN: https://ssrn.com/abstract=2796221*

Brañas-Garza, P., Espín, A. M., Exadaktylos, F., & Herrmann, B. (2014). Fair and unfair punishers coexist in the ultimatum game. *Scientific Reports*, 4, 6025.

Brekke, K. A., Kverndokk, S., & Nyborg, K. (2003). An Economic Model of Moral Motivation. *Journal of Public Economics*, 87, 1967-1983.

Brenner, T., & Vriend, N. J. (2006). On the behavior of proposer in ultimatum games. *Journal of Economic Behavior and Organization*, 61, 617-631.

Camerer, C. F. *Behavioral game theory: Experiments in strategic interaction* Princeton, NJ: Princeton University Press (2003).

Camerer, C. F., & Fehr, E. Measuring social norms and preferences using experimental games: A guide for social scientists. In J. Henrich, R. Boyd, S. Bowles, C. Camerer, E. Fehr, & H. Gintis (Eds.), *Foundations of human sociality* (pp. 55-95). New York, NY: Oxford University Press (2004).

Capraro, V. (2013). A model of human cooperation in social dilemmas. *PLoS ONE*, 8, e72427.

Capraro, V. Corgnet, B., Espín, A. M., & Hernán-González, R. (2017). Deliberation favours social efficiency by making people disregard their relative shares: Evidence from US and India. *Royal Society Open Science*, 4, 160605.

Capraro, V., Jordan, J. J., & Rand, D. G. (2014). Heuristics guide the implementation of social preferences in one-shot Prisoner's Dilemma experiments. *Scientific Reports*, 4, 6790.

Capraro, V., Polukarov, M., Venanzi, M., & Jennings, N. R. (2015). Cooperative equilibrium beyond social dilemmas: Pareto solvable games. *Available at https://arxiv.org/abs/1509.07599*.

Capraro, V., Smyth, C., Mylona, K., & Niblo, G. A. (2014). Benevolent characteristics promote cooperative behaviour among humans. *PLoS ONE*, 9, e102881.

Capraro, V., & Rand, D. G. (2018). Do the right thing: Experimental evidence that preferences for moral behavior, rather than equity or efficiency per se, drive human prosociality. *Judgment and Decision Making*, 13, 99-111.

Chen, Y.-H., Chen, Y.-C., Kuo, W.-J., Kan, K., Yang, C. C., & Yen, N.-S. (2017). Strategic motives drive proposers to offer fairly in Ultimatum Games: An fMRI Study. *Scientific Reports*, 7, 527.

Cooper, D., & Dutcher, G. (2011). The dynamics of responder behavior in ultimatum games: A meta-study. *Experimental Economics*, 14, 519-546.

Corgnet, B., Espín, A. M., &Hernán-González, R. (2015). The cognitive basis of social behavior: cognitive reflection overrides antisocial but not always prosocial motives. *Frontiers in Behavioral Neuroscience*, 9, 287.

Crockett, M. J., Clark, L., Tabibnia, G., Lieberman, M. D., & Robbins, T. W. (2008). Serotonin modulates behavioral reactions to unfairness. *Science*, 320, 1739.

Charness, G., & Rabin, M. (2002). Understanding social preferences with simple tests. *The Quarterly Journal of Economics*, 117, 817-869.

DellaVigna, S., List, J. A., & Malmendier, U. (2012). Testing for altruism and social pressure in charitable giving. *The Quarterly Journal of Economics*, 127, 1-56.

Engelmann, D., & Strobel, M. (2004). Inequality aversion, efficiency, and maximin preferences in simple distribution experiments. *The American Economic Review*, 94, 857-869.

Eriksson, K., Strimling, P., Anderson, P. A., & Lindholm, T. (2017). Costly punishment in the ultimatum game evokes moral concern, in particular when framed as payoff reduction. *Journal of Experimental Social Psychology*, 69, 59-64.

Espín, A. M., Exadaktylos, F., Herrmann, B., & Brañas-Garza, P. (2015). Short- and long-run goals in ultimatum bargaining: Impatience predicts spite-based behavior. *Frontiers in Behavioral Neuroscience*, 9, 214.

Fehr, E., & Fischbacher, U. (2003). The nature of human altruism. *Nature*, 425, 785-781.

Fehr, E. & Schmidt, K. M. (1999). A theory of fairness, competition, and cooperation. *Quarterly Journal of Economics*, 114, 817-868.

Fehr, E., Schmidt, K. M. (2006). The Economics of Fairness, Reciprocity and Altruism -Experimental Evidence and New Theories. In: Kolm, S.C. (ed.): *Handbook of the Economics of Giving, Altruism and Reciprocity. Handbooks in Economics* 23, Vol. 1. Amsterdam, NE: Elsevier.

Forsythe, R., Horowitz, J. L., Savin, N. E., & Sefton, M. Fairness in simple bargaining experiments. *Games and Economic Behavior*, 6, 347-369 (1994).

Grimm, V., & Mengel, F. (2011). Let me sleep on it: Delay reduces rejection rates in ultimatum games. *Economic Letters*, 111, 113-115.

Güth, W., & Kocher, M. G. (2014). More than thirty years of ultimatum bargaining experiments: Motives, variations, and a survey of the recent literature. *Journal of Economic Behavior and Organization*, 108, 396-409.

Güth, W., Schmittberger, R., & Schwarze, B. (1982). An experimental analysis of ultimatum bargaining. *Journal of Economic Behavior and Organization*, 3, 367-388.

Güth, W., Schmidt, C., Sutter, M., 2003. Fairness in the mail and opportunism in the Internet: a newspaper experiment on ultimatum bargaining. *Ger. Econ. Rev.* 4, 243–265.

Henrich, J., Boyd, R., Bowles, S., Camerer, C., Fehr, E., Gintis, H., & McElreath, R. (2001). In search of Homo Economicus: Behavioral experiments in 15 small-scale societies. *The American Economic Review*, 91, 73-78.

Horton, J. J., Rand, D. G., & Zeckhauser, R. J. (2011). The online laboratory: Conducting experiments in a real labor market. *Experimental Economics*, 14, 399-425.

Huck, S., Kübler, D., & Weibull, J. (2012). Social norms and economic incentives in firms. *Journal of Economic Behavior and Organization*, 83, 173-185.

Kessler, J. B., & Leider, S. (2012). Norms and contracting. *Management Science*, 58, 62-77.

Kimbrough, E. O., & Vostroknutov, A. (2016). Norms make preferences social. *Journal of the European Economic Association*, 608-638.

Krupka, E. L., & Weber, R. A. (2013). Identifying social norms using coordination games: Why does dictator game sharing vary? *Journal of the European Economic Association*, 11, 495-524.

Lazear, E. P., & Malmendier, U., & Weber, R. A. (2012). Sorting in experiments with application to social preferences. *American Economic Journal: Applied Economics*, 4, 136-163.

Levitt, S. D., & List, J. A. (2007). What do laboratory experiments measuring social preferences reveal about the real world? *Journal of Economic Perspectives*, 21, 153-174.

Lin, H., & Sunder, S. (2002). Using Experimental Data to Model Bargaining Behavior in Ultimatum Games. In R. Z. a. A. Rapoport (Ed.), *Experimental Business Research* (pp. 373-397). Dordrecht: Kluwer.

Mason, W., & Suri, S. (2012). Conducting behavioral research on Amazon's Mechanical Turk. *Behavior Research Methods*, 44, 1-23.

Messick, D. M., & McClintock, C. G. (1968). Motivational bases of choice in experimental games. *Journal of Experimental Social Psychology*, 4, 1-25.

Nowak, M. A., Page, K., M. & Sigmund, K. (2000). Fairness versus reason in the ultimatum game. *Science*, 289, 1773-1775.

Paolacci, G., & Chandler, J. (2014). Inside the Turk: Understanding Mechanical Turk as a participant pool. *Current Directions in Psychological Science*, 23, 184-188.

Paolacci, G., Chandler, J., & Ipeirotis, P. G. (2010). Running experiments on Amazon Mechanical Turk. *Judgment and Decision Making*, 5, 411-419.

Peysakhovich, A., Nowak, M. A, & Rand, D. G. (2014). Humans display a "cooperative phenotype" that is domain general and temporally stable. *Nature Communications*, 5, 4939.

Rand, D. G. (2012). The promise of Mechanical Turk: How online labor markets can help theorists run behavioral experiments. *Journal of Theoretical Biology*, 299, 172-179.

Rand, D. G., Tarnita, C. E., Ohtsuki, H., & Nowak, M. A. (2013). Evolution of fairness in the one-shot anonymous ultimatum game. *Proceedings of the National Academy of Sciences*, 2581-2586.

Reigstad, A. G., Strømland, E. A., & Tinghög, G. (2017). Extending the cooperative phenotype: Assessing the stability of cooperation across countries. *Frontiers in Behavioral Neuroscience*, 8, 1990.

Roth, A. E., Prasnikar, V., Okuno-Fujiwara, M., Zamir, S. (1991). Bargaining and market behavior in Jerusalem, Lubljana, Pittsburgh, and Tokyo: An experimental study. *The American Economic Review*, 81, 1068-1095.

Sanfey, A. G., Rilling, J. K., Aronson, J. A., Nystrom, L. E., Cohen, J. D. The neural basis of economic decision-making in the ultimatum game. *Science*, 300, 1755-1758.

Schuster, S. (2017). A new solution concept for the Ultimatum Game leading to the Golden Ratio. *Scientific Reports*, 7, 5642.
Suleiman, R. (2017). On gamesmen and fairmen: Explaining fairness in non-cooperative barganing games. *Royal Society Open Science*, 5, 171709.
Sutter, M., Kocher, M. C., & Strauß, S. (2003). Bargaining under time pressure in an experimental ultimatum game. *Economics Letters*, 81, 341-347.
Szolnoki, A., Perc, M., & Szabó, G. (2012a). Accuracy in strategy imitations promotes the evolution of fairness in the spatial ultimatum game. *EuroPhysics Letters*, 100, 28005.
Szolnoki, A., Perc, M., & Szabó, G. (2012b). Defense mechanisms of empathetic players in the spatial ultimatum game. *Physical Review Letters*, 109, 078701.
Thomas, K. A., & Clifford, S. (2017). Validity and Mechanical Turk: An assessment of exclusion methods and interactive experiments. *Computers in Human Behavior*, 77, 184-197.
Van Lange, P. A. M., Otten, W., De Bruin, E. M. N., & Joireman, J. A. (1997). Development of prosocial, individualistic, and competitive orientations: Theory and preliminary evidence. *Journal of Personality and Social Psychology*, 73, 733-746.
Van't Wout, M., Kahn, R. S., Sanfey, A. G, & Aleman, A. (2006). Affective state and decision-making in the Ultimatum Game. *Experimental Brain Research*, 169, 564-568.
Weber, J. M., Kopelman, S., & Messick, D. M. (2004). A conceptual review of decision making in social dilemmas: Applying a logic of appropriatness. *Personality and Social Psychology Review*, 8, 281-307.
Wells, J., & Rand, D. G. (2013). Strategic self-interest can explain seemingly "fair" offers in the Ultimatum Game. *Available at* http://ssrn.com/abstract=2136707
Yamagishi, T., Horita, Y., Takagishi, H., Shinada, M., Tanida, S., & Cook, K. S. (2009). The private rejection of unfair offers and emotional commitment. *Proceedings of the National Academy of Sciences*, 106, 11520-11523.
Yamagishi, T., Horita, Y., Mifune, N., Hashimoto, H., Li, Y., Shinada, M., Miura, A., Inukai, K., Takagishi, H., Simunovic, D. (2012). Rejection of unfair offers in the ultimatum game is no evidence of strong reciprocity. *Proceedings of the National Academy of Sciemces*, 109, 20364-20368.
Yamagishi, T., Li, Y., Fermin, A. S. R., Kanai, R., Tagakishi, H., Matsumoto, Y., Kiyonari, T., & Sakagami, M. (2017). Behavioural differences and neural substrates of altruistic and spiteful punishment. *Scientific Reports*, 7, 14654.

**Experimental instructions**

We report the detailed experimental instructions only of the Main Study. The instructions of the Preliminary Study were identical, apart from the following differences: in case of UG1, the question "You are Person A, what amount do you want offer to Person B?" was replaced by the question ""What do you think is the morally right offer?"; in case of UG 2, the question "You are Person B. Please select below your minimum acceptable offer. That is, if the offer that A gives you is below this, you will reject it, and if the offer that A gives you is above or equal to this, you will accept it" was replaced by the question "What do you think it is the morally right minimum offer for you to accept?".

*Ultimatum Game in the role of the proposer*

In this interaction you are matched with another participant.

One of you will be Person A, one of you will be Person B.

Person A starts with $0.10, Person B starts with $0.

First Person A makes a choice, then Person B responds.

1) Person A will make an offer on how to split the $0.10 with Person B.

2) Person B will either accept or reject this offer.

If Person B accepts, then B will get the offered amount and A will keep the rest. If B rejects the offer, then both individuals will get $0.

Here are some questions to ascertain that you understand the rules. Remember that you have to answer all of these questions correctly in order to get the completion code. If you fail any of them, the survey will automatically end and you will not get any payment.

What is the offer that Person A should make in order to equalize his and Person B's gain?

    A) $0
    B) $0.01
    C) $0.02
    D) $0.03
    E) $0.04
    F) $0.05
    G) $0.06
    H) $0.07
    I) $0.08
    J) $0.09
    K) $0.10

What happens if B accepts an offer of $0.02?

- A gets $0.08 and B gets $0.02.
- A gets $0 and B gets $0.

What happens if B rejects an offer of $0.02?

- A gets $0.08 and B gets $0.02.
- A gets $0 and B gets $0.

Congratulations, you successfully answered all the questions. It is now time to make your decisions.

You are Person A, what amount do you want offer to Person B?

A) $0
B) $0.01
C) $0.02
D) $0.03
E) $0.04
F) $0.05
G) $0.06
H) $0.07
I) $0.08
J) $0.09
K) $0.10

*Ultimatum Game in the role of the responder*

Instructions were exactly the same as in the previous case, a part from the very last sentence, which was replaced with: "You are Person B. Please select below your minimum acceptable offer. That is, if the offer that A gives you is below this, you will reject it, and if the offer that A gives you is above or equal to this, you will accept it."

*Trade-Off game in the "give" frame*

You are Player A. You are playing a game with other two players, Player B and Player C. Each of you starts this game with $0.13.

You get to make a choice (Player B and Player C do not make any decisions).

You can either choose to be nice or not. If you choose to be nice, you earn an additional $0.02 and Player B earns an additional $0.10. If you choose not to be nice, no one earns any additional money and you all end the game with $0.13.

This is the only interaction you have with Player B and Player C. They will not have the opportunity to influence your gain in later parts of the HIT.

What do you want to do?

A) Be nice
B) Don't be nice

### *Trade-Off game in the "equalize" frame*

You are playing a game with other two players, Player B and Player C. You start this game with $0.15, Player B starts with $0.23 and Player C with $0.13.

You get to make a choice (Player B and Player C do not make any decisions).

You can either choose to be nice or not. If you choose to be nice, you give up $0.02 to restore equality, so that you all earn $0.13. If you choose not to be nice, no changes are made to the payoffs, and you each earn what you have started with.

This is the only interaction you have with Player B and Player C. They will not have the opportunity to influence your gain in later parts of the HIT.

What do you want to do?

A) Be nice
B) Don't be nice